\documentclass[showpacs,prd,twocolumn,floatfix,
nofootinbib,superscriptaddress]{revtex4}
\usepackage{graphicx}

\newcommand{\be}{\begin{equation}}
\newcommand{\ee}{\end{equation}}
\newcommand{\nl}{\nonumber \\}

\begin{document}


\title{$|V_{cd}|$ from D Meson Leptonic Decays}

\author{Heechang Na} 
\affiliation{Argonne Leadership Computing Facility,
ANL, Argonne, IL 60439, USA.}
\author{Christine T.\ H.\ Davies}
\affiliation{SUPA, School of Physics \& Astronomy,
University of Glasgow, Glasgow, G12 8QQ, UK}
\author{Eduardo Follana}
\affiliation{Departamento de Fisica Teorica, 
Universidad de Zaragoza, E-50009 Zaragoza, Spain} 
\author{{G.\ Peter} Lepage}
\affiliation{Laboratory of Elementary Particle Physics,
Cornell University, Ithaca, NY 14853, USA}
\author{Junko Shigemitsu}
\affiliation{Department of Physics,
The Ohio State University, Columbus, OH 43210, USA}

\collaboration{HPQCD Collaboration}
\noaffiliation
\date{\today}


\begin{abstract}
We present an update of the $D$ meson decay constant $f_D$ using the Highly Improved 
Staggered Quark (HISQ) action for valence charm and light quarks on MILC $N_f = 2+1$ 
lattices.  The new determination incorporates HPQCD's improved scale $r_1^{N_f = 2+1} 
= 0.3133(23)$fm, accurately retuned bare charm quark masses and data from an 
ensemble that is more chiral than in our previous calculations. 
 We find $f_D = 208.3(3.4)$MeV.  Combining the new $f_D$ with  
 $D \rightarrow \mu \nu_\mu$ branching fraction data from CLEO-c, 
 we extract the CKM matrix element $|V_{cd}| = 0.223(10)_{exp.}(4)_{lat.}$.
This value is in excellent agreement with $|V_{cd}|$ from $D$ semileptonic 
decays and from neutrino scattering experiments and has comparable total errors. 
We determine  the 
ratio  between semileptonic form factor and decay constant 
and find $\left [f^{D \rightarrow \pi}_+(0)\, / \, f_D\right ]_{lat.} = 3.20(15)
{\rm GeV}^{-1}$ to be compared with the experimental value of 
 $\left [f^{D \rightarrow \pi}_+(0)\, / \, f_D \right ] _{exp.} = 3.19(18) 
{\rm GeV}^{-1}$. Finally, we mention recent preliminary but already more 
accurate  $D \rightarrow \mu \nu_\mu$ branching fraction measurements from 
BES III and discuss their impact on precision $|V_{cd}|$ determinations 
in the future.

\end{abstract}


\maketitle


\section{Introduction}
Determinations of individual elements of the Cabibbo-Kobayashi-Maskawa (CKM) 
matrix allows for many cross checks and consistency tests of the Standard 
Model. In most cases there are several processes that can be used to 
extract the same CKM matrix element each involving
 very different experimental and theory inputs. 
For the CKM matrix element $|V_{cd}|$, PDG2010 \cite{pdg} quotes values coming from  
$D \rightarrow \pi, l \nu$ semileptonic decays and from neutrino/antineutrino 
scattering.   The HPQCD collaboration recently published a new
 calculation of $|V_{cd}|$ that reduced errors in the 
 semileptonic decay determination by more than a factor of two \cite{dtopi}, making
it competitive with the neutrino scattering result. 
 In the current 
article we present a third, independent determination based 
this time on $D$ meson leptonic decays. We find a value for 
$|V_{cd}|$ in  complete agreement with the other two determinations 
and with comparable total errors.

The branching fraction for the leptonic decay of a charged $D$ or $D_s$ 
meson via a virtual $W$ boson is given to lowest order by,
\be
\label{brfrac}
{\cal B}(D_q \rightarrow l \nu) = \frac{G_F^2}{8 \pi} f_{D_q}^2 m_l^2 M_{D_q} 
\left ( 1 - \frac{m_l^2}{M_{D_q}^2} \right )^2 |V_{cq}|^2,
\ee
where $m_l$ is the charged lepton mass and $q = d$ or $s$. Electromagnetic 
corrections to this formula are known and routinely taken into account 
by experimentalists in their analyses \cite{rstone,dobkron}. 
Equation (\ref{brfrac}) tells us that 
determination of $|V_{cd}|$ from $D$ leptonic decays requires 
theory to provide the $D$ meson decay constant $f_D$ which is a 
pure QCD nonperturbative quantity. The first lattice QCD calculations 
of $f_D$ and $f_{D_s}$ that included sea quark contributions were 
carried out by the Fermilab Lattice \& MILC collaborations \cite{fermimilc1} and this 
predated experimental studies of these decays. Subsequent experimental 
measurements were consistent with the lattice predictions within errors that 
were more substantial then than they are today for both theory and experiment.
  The initial lattice calculations employed an effective theory 
approach (the heavy clover action \cite{hclover}) for the charm quark on the lattice.
In 2007 the HPQCD collaboration introduced the Highly Improved Staggered 
Quark (HISQ) action which represents not only an extremely accurate lattice 
quark action for light quark physics, but also serves as an accurate relativistic 
action for heavier quarks \cite{hisq}. 
The HISQ action has since been used very successfully 
in simulations involving the charm quark such as for charmonium,
 and for $D$ and $D_s$ meson 
decay constants and semileptonic form factors \cite{fdprl,fds2,dtok,dtopi}. 
 In reference \cite{fdprl} HPQCD published the 
first $f_\pi$, $f_K$, $f_D$ and $f_{D_s}$ results from HISQ valence quarks, including 
HISQ charm quarks, on the MILC AsqTad $N_f = 2 + 1$ lattices \cite{milc},
 all with sub 2\% 
errors. At around the same time experimental measurements of $D$ and $D_s$ 
meson leptonic decay branching fractions were improving significantly \cite{cleofd,
cleofds,babarfds}.
And by the middle of 2008 we were facing an interesting situation where 
there was good agreement between experiment and theory
 for $f_D$ but a close to $4\sigma$ discrepancy 
in $f_{D_s}$. Further improvements and scrutiny became crucial. 

The largest systematic error for $f_{D_s}$ in reference \cite{fdprl} came from the 
uncertainty in the scale $r_1$. HPQCD was using an $r_1$ extracted from 
$\Upsilon$ splittings namely $r_1 = 0.321(5)$ fm with $~1.56$\% errors \cite{oldr1}.
In 2010 HPQCD published a much more accurate $r_1$ determination, 
$r_1 = 0.3133(23)$, based 
on several physical quantities and 
an improved continuum extrapolation (from 5 lattice spacings) \cite{r1}.
A change in the scale affects quantities such as $f_{D_s}$ in two ways:
1. the bare strange and charm quark masses must be retuned on each 
ensemble and 2. the conversion from dimensionless decay constant 
(e.g. in units of $r_1$) to the decay constant in physical units 
is modified.   In reference \cite{fds2} HPQCD updated its value for $f_{D_s}$ 
together with $f_\pi$ and $f_K$ using the new $r_1$. Although $f_\pi$ and $f_K$ 
hardly shifted at all upon going from old to new $r_1$, the updated 
$f_{D_s}$ came out about $2.3\sigma$ (3\%) higher than before. As a 
consequence, taking into account also that experimental results were 
changing and moving closer to theory numbers, the discrepancy in 
$f_{D_s}$ between theory and experiment has now shrunk to a
 $1.6 \sigma$ effect. Reference \cite{fds2} did not present a new calculation 
of $f_D$. Instead we took the previous ratio $f_{D_s} / f_D$ from reference 
\cite{fdprl} 
and combined this with the new $f_{D_s}$ to estimate a new $f_D$.

In this article we complete the process of switching to the new $r_1$ scale 
for meson decay constants and present a direct calculation of 
$f_D$ consistently using the new scale.
Since the time of reference \cite{fdprl}  
 experimental errors in the $D \rightarrow \mu, \nu_\mu$ 
branching fraction have improved from $\sim 7.8$\% down to $\sim 4.3$\%  
in the case of CLEO-c \cite{cleofd} 
and new even more accurate measurements are appearing now 
from BES III \cite{bes3}. 
Together with the new $f_D$ of this article 
with its $~\sim1.66$\% error, one can now extract a $|V_{cd}|$ from $D$ meson 
leptonic decays 
that is as accurate as those from semileptonic decays or neutrino 
scattering and that promises to become even more precise in the near future.

\section{ The Lattice Setup }

\begin{table}
\caption{
Simulation details on three ``coarse'' and three ``fine'' MILC ensembles.
}
\begin{tabular}{|c|c|c|c|c|c|}
\hline
Set &  $r_1/a$ & $m_l/m_s$ (sea)   &  $N_{conf}$&
 $N_{tsrc}$ & $L^3 \times N_t$ \\
\hline
\hline
C1  & 2.647 & 0.005/0.050   & 1200  &  2 & $24^3 \times 64$ \\
\hline
C2  & 2.618 & 0.010/0.050  & 1200   & 2 & $20^3 \times 64$ \\
\hline
C3  & 2.644 & 0.020/0.050  &  600  & 2 & $20^3 \times 64$ \\
\hline
\hline
 F0  & 3.695  &  0.0031/0.031  & 600  & 4 & $40^3 \times 96$ \\
\hline
F1  & 3.699 & 0.0062/0.031  & 1200  & 4  & $28^3 \times 96$ \\
\hline
F2  & 3.712 & 0.0124/0.031  & 600  & 4 & $28^3 \times 96$ \\
\hline
\end{tabular}
\end{table}

\begin{table}
\caption{
Valence quark masses
}
\begin{tabular}{|c|c|c|c|}
\hline
Set & $a m_l$  & $a m_s$ & $a m_c$   \\
\hline
\hline
C1  &  0.0070 &  0.0489 & 0.6207    \\
\hline
C2  & 0.0123 & 0.0492  &  0.6300   \\
\hline
C3  & 0.0246 & 0.0491 & 0.6235   \\
\hline
\hline
 F0  &  0.00339  & 0.0339 & 0.4130   \\
\hline
F1  & 0.00674 & 0.0337 & 0.4130   \\
\hline
F2  & 0.0135  & 0.0336 & 0.4120   \\
\hline
\end{tabular}
\end{table}

Table I lists the three coarse ($a \approx 0.12$fm) and 
three fine ($a \approx 0.09$fm) MILC ensembles used in this study 
together with some lattice details. And in Table II
we show the values for valence quark masses.
 For $f_D$ we have focused more on ensuring 
better control over chiral extrapolations by adding a more chiral fine 
ensemble (Set F0) rather than going to finer lattices as we did in 
reference \cite{fds2} for $f_{D_s}$.
The bare charm quark mass is tuned using the physical $\eta_c$ mass adjusted for 
the absence of electromagnetic, charm sea and annihilation contributions 
in our simulations which leads  to a target value of
$M^{target}_{\eta_c}= 2.985(3)$ GeV \cite{gregory} 
rather than the experimental value of $M_{\eta_c}^{exp}=2.980(1)$ GeV. 
Most of the charm quark mass tuning had been done already in reference \cite{dtok}
for our $D \rightarrow K, l \nu$ studies.
For the present calculations we needed to add tunings only on ensemble F0.
  Figure 1. shows the tuned $\eta_c$ masses for all 6 ensembles.  The bulk of the 
errors shown comes from the $\sim0.1$\% uncertainty in $r_1/a$, whereas the tiny 
black error bars represent the statistical errors on each data point.
A similar plot for tuning of the strange quark mass via the $\eta_s$ 
(fictitious) meson mass is given in Figure 3 of reference \cite{dtok}. 
 And in references \cite{fdprl, fds2, dtok}  
we have demonstrated that once quark masses have been fixed by $\eta_c$ and 
$\eta_s$ then masses for the $D$ and $D_s$ mesons can be derived with 
zero adjustable parameters in good agreement with experiment.  We do not 
repeat those calculations here. However, since we have new data for the 
mass difference $\Delta M_D \equiv M_{D_s} - M_D$, we summarize them in 
an Appendix and compare with $\Delta M_B \equiv M_{B_s} - M_B$
in the $B$ system taken from reference \cite{fbfbs}. 

Having fixed the quark masses we evaluated $D$ and $D_s$ correlators 
on each of the 6 ensembles.  We use random wall sources with 
a different set for each color component
in order to improve statistical errors.
 In the next section we describe how we extract meson decay constants 
from these correlators.

\begin{figure}
\includegraphics*[width=7.0cm,height=9.0cm,angle=270]{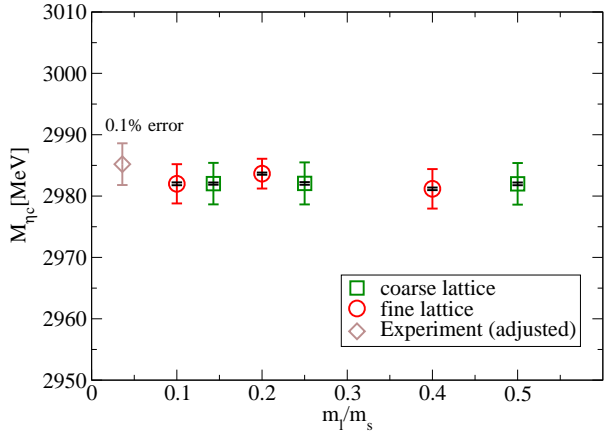}
\caption{
 Checking the tuning of the charm quark mass to the $\eta_c$ meson mass. 
Errors on the simulation results include statistical (black error bars) plus 
errors arising from the uncertainty in $r_1/a$ for each ensemble. 
The ``experimental'' $\eta_c$ mass has 
been adjusted to take into account the lack of annihilation and 
electromagnetic effects in our lattice calculation.
 }
\end{figure}

\section{ Correlators and Fitting Strategies}

The decay constant $f_{D}$ of a pseudoscalar meson made out of a charm quark 
and a light antiquark of mass $m_q$ is defined in terms of the 
matrix element of the heavy-light axial vector current $A_\mu = \overline{\Psi}_q 
\gamma_\mu \gamma_5 \Psi_c$ between the hadronic vacuum and the $D$ meson state.
\be
\langle 0 | A_\mu | D \rangle = p_\mu f_{D}
\ee
Since we employ the relativistic HISQ action for all quarks we are able to 
take advantage of PCAC, as is routinely done for $f_\pi$ and $f_K$, and 
express the decay constant in terms of the pseudoscalar density $PS = \overline{\Psi}
_q \gamma_5 \Psi_c$,
\be
\label{fdq}
f_D = \frac{m_c + m_q}{M_{D}^2} \langle 0 | PS | D \rangle. 
\ee
The relevant hadronic matrix element in eq.(\ref{fdq}) can be extracted from the 
$D$ meson two-point correlator,
\be
\label{c2pnt}
C_{D}^{2pnt}(t) = 
\frac{1}{L^3}\sum_{\vec{x}} \sum_{\vec{y}} \langle 0| \Phi_D(\vec{y},t) 
\Phi_D^\dagger(\vec{x},0)|0 \rangle,
\ee
where $\Phi_D \equiv PS \times a^3$ is the same as the pseudoscalar density 
in lattice units, and is used 
here also as an interpolating operator for the $D$ meson. 
We fit $C_D^{2pnt}(t)$ to the form,
\begin{eqnarray}
\label{twopntfit}
C^{2pnt}_{D}(t) &=& \sum_{k=0}^{N_D-1} |b^D_k|^2 (e^{-E^D_k t} + e^{-E^D_k ( N_t - t)}) \nl
&+& \sum_{k=0}^{N_D^\prime - 1} |d^D_k|^2 (-1)^t 
( e^{-E^{\prime D}_k t} + e^{-E_k^{\prime D} (N_t - t)}).\nl
\end{eqnarray}
The ground state amplitude $b_0^D$ is related to the matrix element of interest 
as,
\be
\label{b0d}
|b^D_0|^2 \equiv \frac{|\langle 0| \Phi_D | D \rangle|^2}{2 M_{D} a^3}
= \frac{|\langle 0| PS | D \rangle|^2 a^3}{2 M_{D}},
\ee
and hence,
\be
\label{afdq}
a f_{D} = \frac{m_c + m_q}{M_{D}} \sqrt{\frac{2}{ a M_{D}}} \, |b^D_0|.
\ee
Our initial goal is to extract the amplitude $b_0^D \equiv |b_0^D|$ as accurately as possible.  
In references \cite{dtok,dtopi}
 we found that fit results for two-point energies and amplitudes 
are improved significantly if one carries out simultaneous fits to two-point 
and three-point correlators. Three-point correlators are calculated, for instance, 
when one studies $D \rightarrow \pi, l \nu$ semileptonic decays. For pions at 
zero momentum one has,
\begin{eqnarray}
\label{thrpnt}
& & C^{3pnt}_{D \rightarrow \pi}(t,T) = 
\frac{1}{L^3} \sum_{\vec{x}}
\sum_{\vec{y}} \sum_{\vec{z}}  \nl
&& \qquad \langle \Phi_D(\vec{y},T) \,\tilde{S}(\vec{z},t) \, 
\Phi^\dagger_\pi(\vec{x},0)
 \rangle,
\end{eqnarray}
where $\tilde{S}$ is the heavy-light scalar density $\overline{\Psi}_c \Psi_q$ 
in lattice units. $C^{3pnt}_{D \rightarrow \pi}$ must be fit to the form,
\begin{eqnarray}
\label{thrpntfit}
& &C_{D \rightarrow \pi}^{3pnt} (t,T)
 = \sum_j^{N_\pi-1} \sum_k^{N_D-1} A_{jk} e^{-E_j^\pi t} e^{ -E_k^D (T-t)} \nl
& & \quad + \sum_j^{N_\pi-1} \sum_k^{N_D^\prime - 1} B_{jk}
 e^{-E_j^\pi t} e^{ -E_k^{\prime D} (T-t)} (-1)^{(T-t)} \nl
& & \quad + \sum_j^{N_\pi^\prime - 1} \sum_k^{N_D-1} C_{jk}
 e^{-E_j^{\prime \pi} t} e^{ -E_k^D (T-t)} (-1)^t \nl
& & \quad + \sum_j^{N_\pi^\prime - 1} \sum_k^{N_D^\prime - 1} D_{jk} e^{-E_j^{\prime \pi} t}
 e^{ -E_k^{\prime D} (T-t)} (-1)^t (-1)^{(T-t)}. \nl
\end{eqnarray}
We will only consider the region $0 \leq t \leq T$ and take $T << N_t$ so that 
any contributions from mesons propagating ``around the lattice'' due to periodic
boundary conditions in time can be ignored.
 The same energies $E^D_k$ and $E^{\prime D}_k$ appear 
in (\ref{twopntfit}) and (\ref{thrpntfit}). Doing simultaneous fits to 
$C^{2pnt}_D$ and $C^{3pnt}_{D \rightarrow \pi}$ places tighter constraints on 
these energies and this helps in reducing fitting errors in the two-point 
amplitudes $b_k^D$. In this way the three-point correlator is acting like 
a very complicated but effective smearing for the propagation of $D$ mesons. 
Normally this would also be considered a very expensive smearing, however we already 
had simulation results for $C^{3pnt}_{D \rightarrow \pi}$ on five out of the six 
ensembles in Table I from the $D$ semileptonic 
project published in reference \cite{dtopi} so we could take advantage of this. 
It was only necessary to create new three-point correlator data on ensemble F0 
 and this only for zero momentum pions.

\begin{figure}
\includegraphics*[width=7.0cm,height=9.0cm,angle=270]{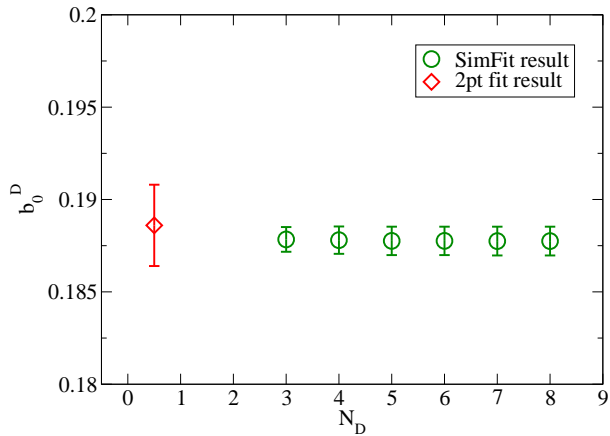}
\caption{Comparison of fit results for the amplitude $b^D_0$ for ensemble 
C1 coming from just the two-point correlator (red data point to the left of 
figure) versus results from simultaneous fits to two-point and three-point correlators.
 }
\end{figure}

In Fig.2 we show some results for $b_0^D$ on ensemble C1 versus the 
number of exponentials from simultaneous fits (we set $N_D = N^\prime_D 
= N_\pi = N^\prime_\pi$) and compare with fit results to just $C^{2pnt}_D$ alone. 
One sees the improvement in the fitting errors coming from the simultaneous fits. 
All our fits are done using Bayesian methods \cite{bayse}. 
 We use the ``sequential method'', 
where starting from $N =$2 or 3 the output from an $N$ - exponential fit becomes 
the initial values for the subsequent $(N +1)$ - exponential fit.

In addition to the $D$ meson decay constant $f_D$ we have also accumulated 
new data for $f_{D_s}$ by studying $D_s$ meson two-point correlators.  
Here we do not have $D_s$ semileptonic decay three-point correlator data. 
So, our extraction of the relevant amplitude $b_0^{D_s}$ was carried out from 
just the two-point correlators.  Since statistical errors are smaller for 
$D_s$ than for $D$ mesons, this lack of ability to carry out simultaneous 
fits in the case of $D_s$ was not a serious problem.  In Table III we list all 
our fit results for $aM_D$, $af_D$, $aM_{D_s}$, $af_{D_s}$ and 
the ratio $f_{D_s}/f_D$.  

\begin{table}
\caption{
Fit Results }
\begin{center}
\begin{tabular}{|c|c|c|c|c|c|}
\hline
 Set  & $aM_D$ &  $af_D$ & $aM_{D_s}$ & $af_{D_s}$ &  $f_{D_s} / f_D$  \\
\hline
\hline
C1  &1.1395(7)&0.1370(5)&1.1878(3)&0.1541(3)&1.1245(37) \\
C2 &1.1591(7)&0.1421(4)&1.2014(4)&0.1566(3)&1.1018(32) \\
C3 &1.1618(5)&0.1464(3)&1.1897(3)&0.1552(3)&1.0600(16) \\
\hline
F0 &0.8096(3)&0.0943(2)&0.8471(1)&0.1074(1)&1.1385(22) \\
F1 &0.8130(3)&0.0966(2)&0.8471(2)&0.1082(1)&1.1202(21)  \\
F2 &0.8189(3)&0.1001(2)&0.8434(2)&0.1076(1)&1.0750(14) \\
\hline
\end{tabular}
\end{center}
\end{table}

\section{ Chiral and Continuum Extrapolation }

The next goal is to extrapolate the entries for $f_D$ in Table III to the 
continuum and chiral limit. The latter is defined as the limit 
$m_q/m_s \rightarrow 1/27.4$, or using $m_s/m_c = 1/11.85$ from reference 
\cite{msmcrat}, the 
limit $m_q/m_c \rightarrow 1/(27.4 \times 11.85)$. We carry out the simultaneous 
chiral/continuum extrapolation using continuum partially quenched heavy meson chiral 
perturbation theory (PQHMChPT) \cite{sharpe,cacb,fermimilc} 
 augmented by lattice spacing dependent terms.  
This is the same formalism employed recently in our $f_B$ and $f_{B_s}$ 
determinations \cite{fbfbs}.  We write,
\be
\label{fdansatz}
f_D = A ( 1 + \delta f + [analytic] ) ( 1 + [discret.]).
\ee
The chiral logarithm term $\delta f$ is taken from the original literature 
on PQHMChPT \cite{cacb,fermimilc} and is also summarized in the Appendix of
 \cite{fbfbs}. As in that reference 
we take,
\begin{eqnarray}
\label{analytic}
& & [analytic] = \nl
& &\beta_0 (2 m_u + m_s) / \tilde{m}_c + \beta_1 m_q/m_c 
+ \beta_2 (m_q/m_c)^2, \nl
\end{eqnarray}
where $m_u (m_q)$ is the sea (valence) light quark mass. $\tilde{m}_c$ is the 
AsqTad charm quark mass tuned to the $\eta_c$ meson made out of AsqTad 
charm quark and antiquark, and is the appropriate charm quark mass to use 
for sea quarks.
  We take $\tilde{m}_c$  from reference \cite{hisq} where 
it was found that $\tilde{m}_c /m_c \approx 0.9$ for lattices employed  
in the current article.
 Using ratios of 
bare quark masses to parameterize the ``$analytic$'' terms is convenient 
since such ratios are scale independent.
 We use the valence charm quark mass as the scale to measure 
the dominant discretization effects and set,
\be
\label{discret}
[discret.] = c_0 (a m_c)^2 + c_1 (am_c)^4.
\ee
We will call the chiral/continuum extrapolation ansatz given by 
eq.(\ref{fdansatz}) together with (\ref{analytic}), 
(\ref{discret}) and eq.(A7) of reference \cite{fbfbs} for $\delta f$ our 
``basic ansatz''.  The result of the extrapolation to the 
physical point using the basic ansatz is given by the green square 
point in Fig.3.  We have tested the stability of this result by 
modifying the basic ansatz in a number of ways and redoing the 
extrapolation.  The modifications that were tried out are the following :
\begin{enumerate}
\item dropping the $\beta_2$ term in (\ref{analytic})
\item adding a $(m_q/m_c)^3$ term in (\ref{analytic})
\item dropping the $c_1$ term in (\ref{discret})
\item adding $(am_c)^n$, n = 6, 8, 10, to (\ref{discret})
\item replacing $c_i$ in (\ref{discret}) by $ c_i \times$[power series in 
$(m_q/m_c)$]
\item using powers of $(a/r_1)$ rather than of $(a m_c)$ in (\ref{discret})
\item using eq.(A1) of reference \cite{fbfbs} 
 rather than (A7) for the chiral logarithm term $\delta f$
\item allowing for a 20\% error in $f_\pi$ entering the chiral perturbation theory 
formulas
\end{enumerate}
Fig.4 compares the extrapolation results with these modifications 
in place with the basic ansatz value at the physical point.

\begin{figure}
\includegraphics*[width=7.0cm,height=9.0cm,angle=270]{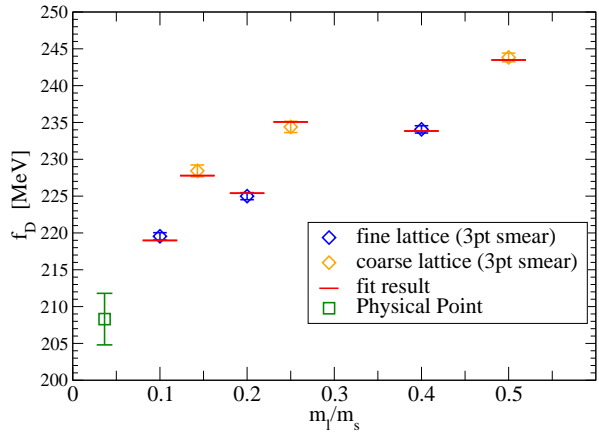}
\caption{Results for $f_D$ versus $m_l/m_s$ and at the physical point
 }
\end{figure}

\begin{figure}
\includegraphics*[width=7.0cm,height=9.0cm,angle=270]{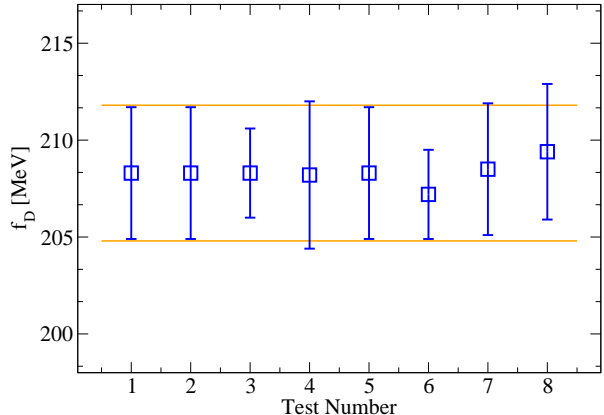}
\caption{Testing the stability of the chiral/continuum extrapolation.
The numbers on the horizontal axis refer to the type of modification 
applied to the basic ansatz of  (\ref{fdansatz}), (\ref{analytic}) and 
(\ref{discret}) as listed in the text. The orange horizontal lines 
bracket the basic ansatz result, i.e. the physical point result  
in Fig.3.
 }
\end{figure}

\section{Results}

\begin{table}
\caption{
Error budget }
\begin{center}
\begin{tabular}{|c|c|c|c|}
\hline
 Source  & $f_D$ &  $f_{D_s}$ &  $f_{D_s} / f_D$  \\
        &  (\%)  &  (\%) & (\%)  \\
\hline
\hline
statistics/fitting   &  0.5 &  0.3 & 0.3\\
scale $r_1$  &  0.7 & 0.7&  ---\\
$r_1/a$  &    0.04  & 0.05 &  ---\\
continuum extrap. &  1.2 & 1.2 &  0.9\\
chiral extrap. \& $g_{D^*D\pi}$  &  0.7 & 0.2&  0.5 \\
mass tunings & 0.1  & 0.2  & 0.2 \\
finite volume & 0.3  & 0.1  & 0.3 \\
\hline
 Total  &  1.7 &  1.5&  1.1\\
\hline
\end{tabular}
\end{center}
\end{table}

Table IV gives the error budget for $f_D$, $f_{D_s}$ and $f_{D_s}/f_D$.
 For all but the last two entries we use the methods of reference \cite{alphas} to 
isolate contributions from different sources that 
make up  the total error coming out of the 
chiral/continuum extrapolations. For the finite volume error 
we take over the result from reference \cite{fdprl} where an analysis was carried out 
comparing finite and infinite volume chiral perturbation theory.

Taking all errors into account our final value for $f_D$ is,
\be
\label{fdval}
f_D = 208.3(1.0)_{stat.}(3.3)_{sys.} {\rm MeV}.
\ee
This is in good agreement with the previous result of $f_D=207(4)$MeV 
\cite{fdprl} based 
on HPQCD's old $r_1$, but is slightly more accurate. Eq.(\ref{fdval}) represents 
 the most precise $f_D$ available today.

For completeness we also give new 
values for $f_{D_s}$ and $f_{D_s}/f_D$,
\be
\label{fds}
             f_{D_s} = 246.0(0.7)_{stat.} (3.5)_{sys.} {\rm MeV},
\ee
and
\be
f_{D_s}/f_D = 1.187(4)_{stat.}(12)_{sys.}.
\ee
The result for $f_{D_s}$, eq.(\ref{fds}), is consistent with HPQCD's 
best updated value of $f_{D_s} = 248.0(2.5)$MeV \cite{fds2} but is not as accurate. 
One sees from Table IV that the dominant error 
comes from the continuum extrapolation. In this respect the current calculation 
of $f_{D_s}$ 
is not competitive with reference \cite{fds2} 
 which employed data from five lattice spacings.

The new $f_D$ of eq.(\ref{fdval}) can be combined with the $D \rightarrow 
\mu, \nu_\mu$ branching fraction from CLEO-c \cite{cleofd} 
 to extract a new value for $|V_{cd}|$. 
We find,
\be
\label{vcd}
|V_{cd}|_{lepton. d.} = 0.223(10)_{exp.}(4)_{lat.}.
\ee
The first error, which is the experimental error, dominates the total 
error of 4.8\%. Eq.(\ref{vcd}) agrees very well with HPQCD's recent 
determination of $|V_{cd}|$ from $D \rightarrow \pi, l \nu$ semileptonic 
decays \cite{dtopi}, 
namely $|V_{cd}|_{semilep. d.} = 0.225(6)_{exp.}(10)_{lat.}$, where now the 
lattice error dominates over the one from experiment.  Both leptonic and 
semileptonic determinations agree with $|V_{cd}| = 0.230(11)$ \cite{pdg} coming from 
neutrino scattering, and all three have comparable total errors.

As mentioned in the Introduction, BES III has recently announced 
preliminary results for the $D \rightarrow \mu \nu_\mu$ 
branching fraction \cite{bes3}. Using their numbers we find,
\be
\label{besvcd}
|V_{cd}|^{BES III}_{lepton. d.} = 0.220(7)_{exp.}(4)_{lat.} \quad 
{\rm [preliminary]}.
\ee
which agrees well with (\ref{vcd}) and has smaller experimental errors.

Another way to check the consistency of the Standard Model and/or to 
test the lattice approach to heavy flavor physics is to consider the ratio between 
semileptonic form factor and decay constant $f_+^{D \rightarrow \pi}(0)/f_D$.
We find, by combining eq.(\ref{fdval}) with $f_+^{D \rightarrow \pi}(0)=0.666(29)$ 
from reference \cite{dtopi},
\be
\label{ratlat}
\left [ f_+^{D \rightarrow \pi}(0) \, / \, f_D \right ]_{lat.} = 3.20(15)
{\rm GeV}^{-1}.
\ee
This can be compared with the experimental ratio in which $|V_{cd}|$ cancels of 
\cite{cleofd,cleoc2}
\be
\label{ratexp}
\left [ f_+^{D \rightarrow \pi}(0) \, / \, f_D \right ]_{exp.} = 3.19(18)
{\rm GeV}^{-1}.
\ee
Eq.(\ref{fdval}), eq.(\ref{vcd}) and the good agreement between (\ref{ratlat}) and 
(\ref{ratexp}) are the main results of this article.

\section{Summary}
In this article we presented a new determination of the CKM matrix element 
$|V_{cd}|$, eq.(\ref{vcd}), 
 made possible by an updated calculation of the decay constant $f_D$, 
eq.(\ref{fdval}),
and improved determinations of the $D \rightarrow \mu, \nu_\mu$ 
leptonic decay branching fraction by CLEO-c \cite{cleofd} and BES III 
\cite{bes3}. 
In Fig.5 we compare the new $f_D$ with HPQCD's previous value \cite{fdprl} and 
with results from other lattice collaborations \cite{fermimilc,etmc,pacs}.  
And in Fig.6 we plot different 
results for $|V_{cd}|$ including the leptonic decay determination of this article, 
together with semileptonic decay and neutrino scattering determinations.

\begin{figure}
\includegraphics*[width=8.0cm,height=9.0cm,angle=270]{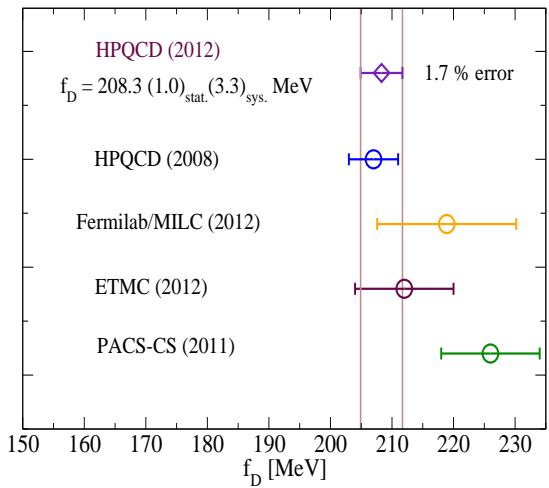}
\caption{Result for $f_D$ from this work and comparisons with 
previous work \cite{fdprl,fermimilc,etmc,pacs}.
 }
\end{figure}

\begin{figure}
\includegraphics*[width=8.0cm,height=9.0cm,angle=270]{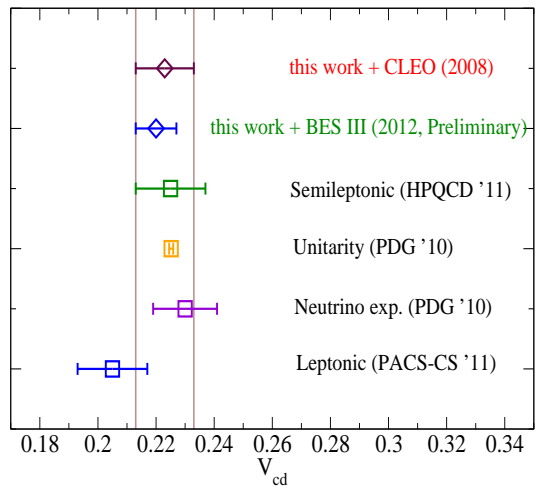}
\caption{Summary of $|V_{cd}|$ determinations from leptonic and 
semileptonic $D$ decays, from neutrino scattering and from unitarity.
 }
\end{figure}

In the future it will be important to continue working on reducing the theory 
errors in eq.(\ref{ratlat}) and the experimental errors in eq.(\ref{ratexp}).  
The former is dominated by errors in the lattice 
determination of $f_+^{D \rightarrow \pi}(0)$ and work is underway to 
significantly reduce them \cite{jonna}.  The experimental error in 
eq.(\ref{ratexp}) comes mainly from the leptonic decay branching fraction and 
one can look forward to improvements there as well. In particular, the recent 
measurements by BES III \cite{bes3} look very promising. 
 The crucial question 
is whether the nice agreement seen now between eq.(\ref{ratlat}) and 
eq.(\ref{ratexp}) 
will continue to hold once errors dip down to $\sim$2\% or below.   

\vspace{.2in}
\noindent
{\bf Acknowledgements}: \\
This work was supported by the DOE (DE-FG02-91ER40690 
and  DE-AC02-06CH11357) and
the NSF (PHY-0757868) in the U.S., by the STFC in the U.K., 
by MICINN (FPA2009-09638 and FPA2008-10732) and DGIID-DGA (2007-E24/2) in Spain, and by ITN-STRONGnet (PITN-GA-2009-238353) in the EU.  
E. Follana is supported on the MICINN Ramon Y Cajal program.
Numerical simulations were
 carried out on facilities
of the USQCD collaboration funded by the Office of Science of the DOE and 
at the Ohio Supercomputer Center.
We thank the MILC collaboration for use of their gauge configurations.

\appendix

\section{The $D_s$ - $D$ Mass Difference}

\begin{table}
\caption{
Mass Splittings in the $D$ and $B$ systems. The $\Delta M_B$ 
numbers are taken from \cite{fbfbs}. }
\begin{center}
\begin{tabular}{|c|c|c|c|}
\hline
Set  & $\Delta M_D$ [MeV]&  $\Delta M_B$ [MeV] &  $\Delta M_D - \Delta M_B$
 [MeV]  \\
\hline
\hline
C1 &80.4(1.1)&64.8(2.2)& 15.6(2.5)  \\
C2 &69.7(1.0)&57.7(1.8)& 12.0(2.1) \\
C3 &46.5(5)&41.3(2.0)&5.2(2.1)  \\
F0 &87.3(7)&71.7(2.9)&15.6(3.0)  \\
F1 &79.4(7)&61.4(2.0)&18.0(2.1)  \\
F2 &57.4(4)&47.8(1.3)&9.6(1.4)  \\
\hline
\end{tabular}
\end{center}
\end{table}

\begin{figure}
\includegraphics*[width=8.0cm,height=9.0cm,angle=270]{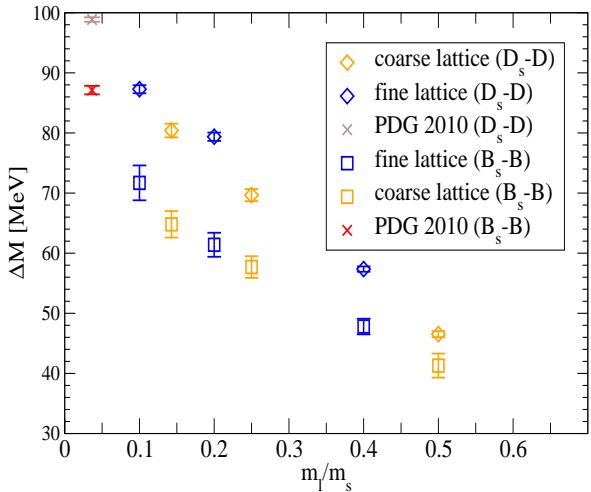}
\caption{Comparison of the mass differences $\Delta M_D = M_{D_s} - M_D$ 
and $\Delta M_B = M_{B_s} - M_B$.
 }
\end{figure}

\begin{figure}
\includegraphics*[width=8.0cm,height=9.0cm,angle=270]{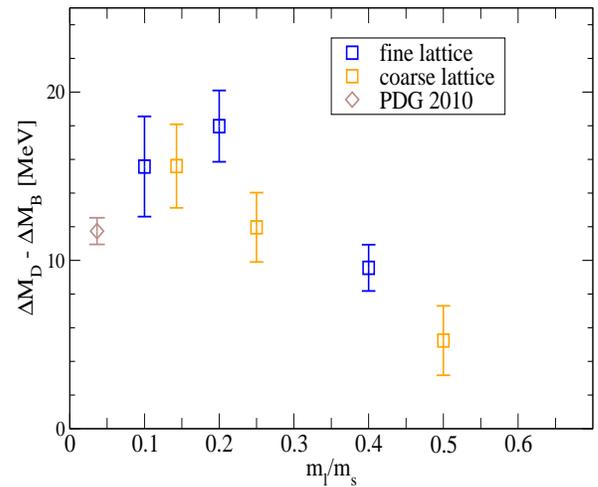}
\caption{$\Delta M_D - \Delta M_B$, the difference of the mass differences 
in the $D$ and $B$ systems.
 }
\end{figure}

In this appendix we summarize results for the mass difference
 $\Delta M_D = M_{D_s} - M_D$ and compare with the analogous difference 
in the $B$ system $\Delta M_B = M_{B_s} - M_B$, where the latter was calculated in
 reference \cite{fbfbs}  employing NRQCD $b$-quarks.
This is an interesting quantitiy to compare since the leading heavy quark 
mass dependence cancels in each of the mass differences and one is testing 
whether the subleading contributions are accurate enough to be able to distinguish 
between the $D$ and $B$ systems. In the difference of differences $\Delta M_D 
- \Delta M_B$  any mistunings of the strange quark mass should also cancel out
 (identical strange and light quark propagators are used in the 
$B$/$B_s$ and the $D$/$D_s$ calculations).    
Table V lists simulation results for $\Delta M_D$, $\Delta M_B$ and for 
$\Delta M_D - \Delta M_B$.  The first two quantities are plotted 
in Fig.7 versus $m_l/m_s$.
  For $\Delta M_D$ statistical errors are small 
enough so that a slight lattice spacing dependence is detected.  Errors are larger 
for $\Delta M_B$ and no discretization effects are visible. Fig.8 shows 
$\Delta M_D - \Delta M_B$.  Ones sees agreement with experiment at the 
1 $\sigma$ level, with a $\sigma$ corresponding to about 5 MeV. 
With current levels of improvements to the lattice actions, 5 MeV appears to be the 
accuracy with which the HISQ action or the combined NRQCD/HISQ actions are 
 able to describe charm-light or bottom-light boundstate dynamics.



\end{document}